\documentclass[11pt, reqno]{amsart}
\usepackage{graphicx,amsmath,amssymb,epsfig}
\usepackage{amsfonts}

\newtheorem{theorem}{Theorem}
\theoremstyle{plain}

\newtheorem{claim}{Claim}

\newtheorem{definition}{Definition}
\newtheorem{example}{Example}

\usepackage{hyperref}

\hypersetup{
    colorlinks=true,
    urlcolor=blue,
    linkcolor=blue,
    citecolor=blue
}

\begin{document}
\title{Identification of Matching Complementarities: A Geometric Viewpoint}
\author{Alfred Galichon{\small $^{\S }$}}
\date{May 29, 2013. This research has received funding from the European
Research Council under the European Union's Seventh Framework Programme
(FP7/2007-2013) / ERC grant agreement no 313699. Support from FiME,
Laboratoire de Finance des March\'{e}s de l'Energie (www.fime-lab.org) is
gratefully acknowledged. Gen Tang provided excellent research assistance.
The author thanks a referee and the editors of this volume for comments that
helped him improving this work.

This is an author-created, un-copyedited version of an article accepted for publication in \textit{Advances in Econometrics, vol. 31: Structural Econometric Models}, DOI: \url{https://doi.org/10.1108/S0731-9053(2013)0000032005}.}
\begin{abstract}
We provide a geometric formulation of the problem of identification of the
matching surplus function and we show how the estimation problem can be
solved by the introduction of a generalized entropy function over the set of
matchings.
\end{abstract}
\maketitle

\noindent


{\footnotesize \textbf{JEL codes}: C78, D61, C13.\vskip50pt }



\bigskip

\section{Setting}

We consider the Becker model of the marriage market as a bipartite matching
game with transferable utility. Let $\mathcal{X}$ and $\mathcal{Y}$ be
finite sets of \textquotedblleft types\textquotedblright\ of men and women
where $\left\vert \mathcal{X}\right\vert =d_{x}$ and $\left\vert \mathcal{Y}%
\right\vert =d_{y}$. Assume that the number of men and women is equal, and
that the number of men of type $x$ (resp. of women of type $y$) is $p_{x}$
(resp. $q_{y}$). We normalize the total number of men and women to one, that
is we set $\sum_{x\in \mathcal{X}}p_{x}=1$ and $\sum_{y\in \mathcal{Y}%
}q_{y}=1$. Let $\Phi _{xy}\geq 0$ be the joint surplus (to be split
endogenously across the pair) from matching a man of type $x$ and a woman of
type $y$. For the clarity of exposition we do not allow for unmatched
individuals.

Recall that under transferable utility, in the Shapley-Shubik model, the
stable matching also maximizes the total surplus
\begin{equation*}
\sum_{x,y}\mu _{xy}\Phi _{xy}
\end{equation*}%
over $\mu \in \mathcal{M}$ the set of \emph{matchings}, defined by%
\begin{equation*}
\mathcal{M}=\left\{ \mu :\mu _{xy}\geq 0,~\sum_{y}\mu
_{xy}=p_{x},~\sum_{x}\mu _{xy}=q_{y}\right\} ,
\end{equation*}%
where $\mu _{xy}$ is interpreted as the number of $\left( x,y\right) $
pairs, which is allowed to be a fractional number.

\bigskip

Note that the equations defining $\mathcal{M}$ have $d_{x}+d_{y}-1$ degrees
of redundancy, hence the dimension of $\mathcal{M}$ is $%
d_{x}d_{y}-d_{x}-d_{y}+1=\left( d_{x}-1\right) \left( d_{y}-1\right) $.

Further, if $\mu $ and $\tilde{\mu}$ are in $\mathcal{M}$, then for $t\in %
\left[ 0,1\right] $, $(t\mu +\left( 1-t\right) \tilde{\mu})$ is also in $%
\mathcal{M}$. Finally, $\mathcal{M}$\ is obviously bounded in $\mathbb{R}%
^{d_{x}d_{y}}$. Hence:

\begin{claim}
The set of matchings $\mathcal{M}$ is a compact convex set of $\mathbb{R}%
^{d_{x}d_{y}}$.
\end{claim}

\bigskip

\section{Identification}

One observes a matching $\hat{\mu}\in \mathcal{M}$ and one wonders whether $%
\hat{\mu}$ is rationalizable, i.e. whether there exists some surplus
function $\Phi $ such that $\hat{\mu}$ is the optimal matching in the
problem with surplus $\Phi $, that is
\begin{equation*}
\hat{\mu}\in \arg \max_{\mu \in \mathcal{M}}\sum_{x,y}\mu _{xy}\Phi _{xy}.
\end{equation*}

As it is classically the case in revealed preference analysis, some
restrictions on $\Phi $ are needed in order to have a meaningful definition.
Indeed, the null surplus function $\Phi _{xy}=0$ always trivially
rationalizes any matching; similarly, $\Phi _{xy}=f_{x}+g_{y}$ which also
rationalizes any $\mu $ as the value of the total surplus evaluated at $\mu $
is $\sum_{x}p_{x}f_{x}+\sum_{y}q_{y}g_{y}$ irrespective of $\mu \in \mathcal{%
M}$. Hence in order to have some empirical bite, we need to impose
\begin{equation*}
\arg \max_{\mu \in \mathcal{M}}\sum_{x,y}\mu _{xy}\hat{\Phi}_{xy}\neq
\mathcal{M}.
\end{equation*}

\bigskip

Let $\mathbf{S}$ be the set of $\Phi $ such that $\Phi _{xy}$ does not
coincides with $f_{x}+g_{y}$ for some vectors $\left( f_{x}\right) $ and $%
\left( g_{y}\right) $. We shall thus seek $\Phi $ in $\mathbf{S}$. The
following assertion characterizes $\Phi $ in dimension two.

\begin{claim}
Assume $d_{x}=d_{y}=2$. Then $\mathbf{S}$ is the set of $\left( \Phi
_{xy}\right) $ such that $\Phi _{11}+\Phi _{22}\neq \Phi _{12}+\Phi _{21}$.
\end{claim}

\bigskip

The previous considerations lead to the following definition:

\begin{definition}
$\hat{\mu}\in \mathcal{M}$ is rationalizable if there is $\hat{\Phi}\in
\mathbf{S}$ such that
\begin{equation}
\hat{\mu}\in \arg \max_{\mu \in \mathcal{M}}\sum_{x,y}\mu _{xy}\hat{\Phi}%
_{xy}.  \label{argmax}
\end{equation}
\end{definition}

Introducing $\mathcal{W}_{0}$ the \emph{indirect surplus function}, defined
as
\begin{equation}
\mathcal{W}_{0}\left( \Phi \right) =\max_{\mu \in \mathcal{M}}\left\langle
\mu ,\Phi \right\rangle  \label{SWF0}
\end{equation}%
where the product $\left\langle \mu ,\Phi \right\rangle $ is defined as%
\begin{equation}
\left\langle \mu ,\Phi \right\rangle =\sum_{xy}\mu _{xy}\Phi _{xy},
\label{scalarProduct}
\end{equation}%
condition (\ref{argmax}) is equivalent, by the Envelope theorem, to
\begin{equation*}
\hat{\mu}\in \partial \mathcal{W}_{0}\left( \hat{\Phi}\right)
\end{equation*}
where $\partial \mathcal{W}_{0}\left( \Phi \right) $ denotes the \emph{%
subgradient} of $\mathcal{W}_{0}$ at $\Phi $. See the Appendix for some
basic results on convex analysis. In the terminology of convex analysis, $%
\mathcal{W}_{0}$ is the \emph{support function} of set $\mathcal{M}$, a
geometric property which we shall develop in the next paragraph.

\bigskip

The following remark is obvious.

\begin{claim}
$\mathcal{W}_{0}$ is positive homogenous of degree one, hence for $t>0$, one
has%
\begin{eqnarray}
\mathcal{W}_{0}\left( t\Phi \right) &=&t\mathcal{W}_{0}\left( \Phi \right)
\label{PH1} \\
\partial \mathcal{W}_{0}\left( t\Phi \right) &=&\partial \mathcal{W}%
_{0}\left( \Phi \right) .  \label{PH0}
\end{eqnarray}
\end{claim}

\section{Geometry}

The following result provides the geometric interpretation of
rationalizability. Formula (\ref{argmax}) means that for $\hat{\mu}$ to be
rationalizable, it needs to maximize a linear functional over the compact
convex set $\mathcal{M}$. As it is well known, a necessary and sufficient
for this to hold is that $\hat{\mu}$ should belong to the boundary of $%
\mathcal{M}$.

\begin{theorem}
The following three conditions are equivalent:

(i) $\hat{\mu}$ is rationalizable,

(ii) $\hat{\mu}$ lies on $\overline{\mathcal{M}}\backslash \mathcal{M}^{int}$%
, the boundary of $\mathcal{M}$,

(iii) There is $\hat{\Phi}\in \mathbf{S}$ such that
\begin{equation}
\hat{\mu}\in \partial \mathcal{W}_{0}\left( \hat{\Phi}\right) .
\label{subgradient}
\end{equation}
\end{theorem}

\begin{figure}[htbp!]
\begin{center}
\includegraphics[width=12cm]{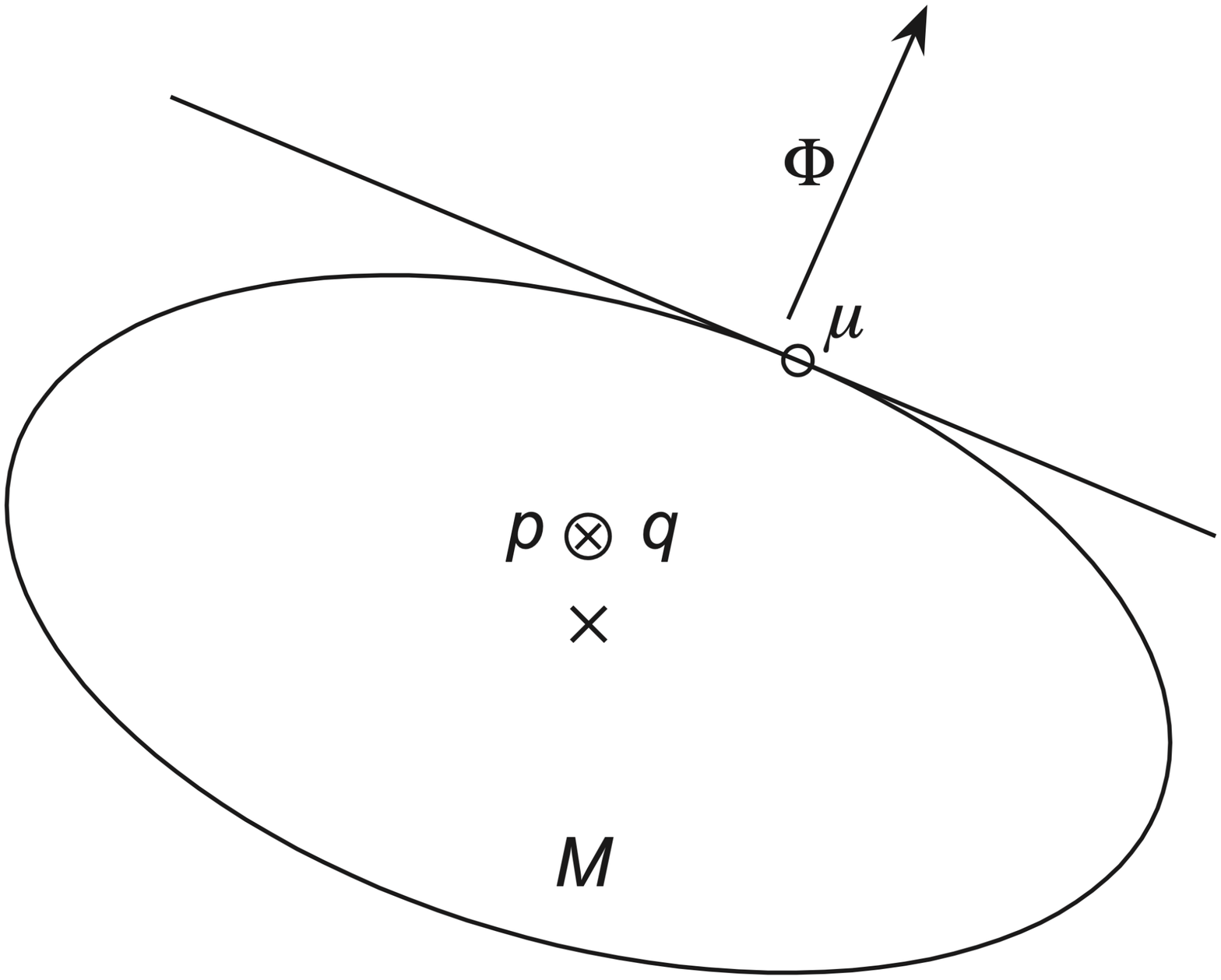}
\caption{Geometric view of rationalizability. In order for matching $\protect\mu $ to be rationalized by surplus function $\Phi $, $\protect\mu $ need to lie on the geometric frontier of $\mathcal{M}$.}
\label{Fig1}
\end{center}
\end{figure}

This theorem is illustrated in Figure \ref{Fig1}. While the equivalence
between part (ii), of geometric kind and part (iii), of analytic nature
follows from standard convex analysis, the insight of this result is to
connect this to the economic notion of rationalizability (i), of revealed
preference flavour. This result provides a geometric understanding of
revealed preference analysis in matching models with transferable utility.
See Echenique et al. (2012).

\bigskip

Geometrically, this means that the matchings that are rationalizable lie on
the boundary of $\mathcal{M}$. We give a very simple example of a $\hat{\mu}$
which is rationalizable.

\begin{example}
Assume $d_{x}=d_{y}=2$ and consider matrix%
\begin{equation*}
\hat{\mu}=\left(
\begin{array}{cc}
1 & 0 \\
0 & 1%
\end{array}%
\right)
\end{equation*}%
then any $\hat{\Phi}$ such that $\hat{\Phi}_{11}+\hat{\Phi}_{22}>\hat{\Phi}%
_{12}+\hat{\Phi}_{21}$ rationalizes $\hat{\mu}$.
\end{example}

We now give a very simple example of a $\hat{\mu}$ which \emph{not} is
rationalizable, i.e. where $\hat{\mu}$ is in the strict interior of $%
\mathcal{M}$.

\begin{example}
Assume $d_{x}=d_{y}=2$ and consider matrix
\begin{equation*}
\hat{\mu}=\left(
\begin{array}{cc}
0.7 & 0.3 \\
0.3 & 0.7%
\end{array}%
\right) .
\end{equation*}%
This matrix is equal to $0.7\left(
\begin{array}{cc}
1 & 0 \\
0 & 1%
\end{array}%
\right) +0.3\left(
\begin{array}{cc}
0 & 1 \\
1 & 0%
\end{array}%
\right) $. Hence for a production function $\Phi $, we get%
\begin{equation*}
\sum_{xy}\hat{\mu}_{xy}\Phi _{xy}=0.7\left( \Phi _{11}+\Phi _{22}\right)
+0.3\left( \Phi _{12}+\Phi _{21}\right)
\end{equation*}%
Hence it cannot be rationalized by a production function $\Phi $ unless $%
\Phi _{11}+\Phi _{22}=\Phi _{12}+\Phi _{21}$. But in that case, set $%
a_{1}=\Phi _{11}$, $b_{1}=0$, $a_{2}=\Phi _{21}$, and $b_{2}=\Phi _{12}-\Phi
_{11}$, thus $\Phi _{ij}=a_{i}+b_{j}$ -- which contradicts $\Phi \in \mathbf{%
S}$. Therefore $\hat{\mu}$ cannot be rationalized.
\end{example}

\begin{example}
As another example, consider $p\otimes q$ defined by $\left( p\otimes
q\right) _{xy}=p_{x}q_{y}$. Clearly, $p\otimes q\in \mathcal{M}$;
intuitively this matching corresponds to matching randomly men and women, so
that the characteristics of the partner are independent. This matching
cannot be rationalized as it lies in the strict interior of $\mathcal{M}$.
Indeed, $p\otimes q$ is the barycenter of the full set $\mathcal{M}$.
\end{example}

\section{Entropy}

In practice, it is almost never the case that a matching $\hat{\mu}$
observed in the population is rationalizable. This is understandable using
the geometric interpretation provided above: the locus of matchings that are
rationalizable being the frontier of a convex set, it is \textquotedblleft
small\textquotedblright\ with respect to the set of matchings that are not
rationalizable, which is the strict interior of this same convex set.

\bigskip

Mathematically speaking, we are looking for a solution $\Phi \in \mathbf{S}$
satisfying%
\begin{equation}
\hat{\mu}\in \partial \mathcal{W}_{0}\left( \Phi \right) .  \label{inclusion}
\end{equation}

If $\mathcal{W}_{0}$ was \textquotedblleft well behaved,\textquotedblright\
more precisely if $\mathcal{W}_{0}$ was strictly convex and continuously
differentiable, then the gradient $\nabla \mathcal{W}_{0}$ would exist and
be invertible with inverse $\nabla \mathcal{W}_{0}^{\ast }$, where $\mathcal{%
W}_{0}^{\ast }$ is the convex conjugate of $\mathcal{W}_{0}$. Then relation (%
\ref{inclusion}) would imply $\Phi =\nabla \mathcal{W}_{0}^{\ast }\left(
\hat{\mu}\right) $. But $\mathcal{W}_{0}$ is not strictly convex, so this
approach does not work, and in fact relation (\ref{inclusion}) has no
solution. Geometrically, it is quite clear why. As remarked above, the image
of $\partial \mathcal{W}_{0}$ is included in the frontier of $\mathcal{M}$,
hence if $\hat{\mu}$ does not lie on the geometric frontier of $\mathcal{M}$%
, then relation (\ref{inclusion}) cannot possibly have a solution.

\bigskip

In order to be able to estimate $\Phi $ based on the observation of $\hat{\mu%
}$, most of the literature following the seminal paper of Choo and Siow
(2005) introduce heterogeneities in matching surpluses. Without trying to be
exhaustive, let us mention Fox (2010, 2011), Galichon and Salani\'{e} (2010,
2012), Decker et al. (2012), Chiappori et al. (2012). As argued in Galichon
and Salani\'{e} (2012), this consists in essence in introducing a \emph{%
generalized entropy function} $\mathcal{I}\left( \mu \right) $ which is
strictly convex, and which is such that%
\begin{equation*}
\mathcal{I}\left( \mu \right) =+\infty \text{ if }\mu \notin \mathcal{M},
\end{equation*}%
such that $\mathcal{I}$ is differentiable on $\mathcal{M}^{int}$ the
interior of $\mathcal{M}$, with, for all $\mu \in \mathcal{M}^{int}$,%
\begin{equation*}
\nabla \mathcal{I}\left( \mu \right) \in \mathbf{S,}
\end{equation*}%
and such $\hat{\Phi}$ is identified by
\begin{equation}
\hat{\Phi}=\nabla \mathcal{I}\left( \hat{\mu}\right) .  \label{FOC}
\end{equation}

Noting that (\ref{FOC}) is the first order condition to the following
optimization program%
\begin{equation}
\mathcal{W}_{\mathcal{I}}\left( \Phi \right) =\max_{\mu \in \mathcal{M}%
}\left\langle \mu ,\Phi \right\rangle -\mathcal{I}\left( \mu \right)
\label{NonHomog}
\end{equation}%
which, as argued in Galichon and Salani\'{e} (2010, 2012), can be
interpreted in some cases as the social welfare of a matching model with
unobserved heterogeneity.

\bigskip

\begin{figure}[htbp!]
\begin{center}
\includegraphics[width=12cm]{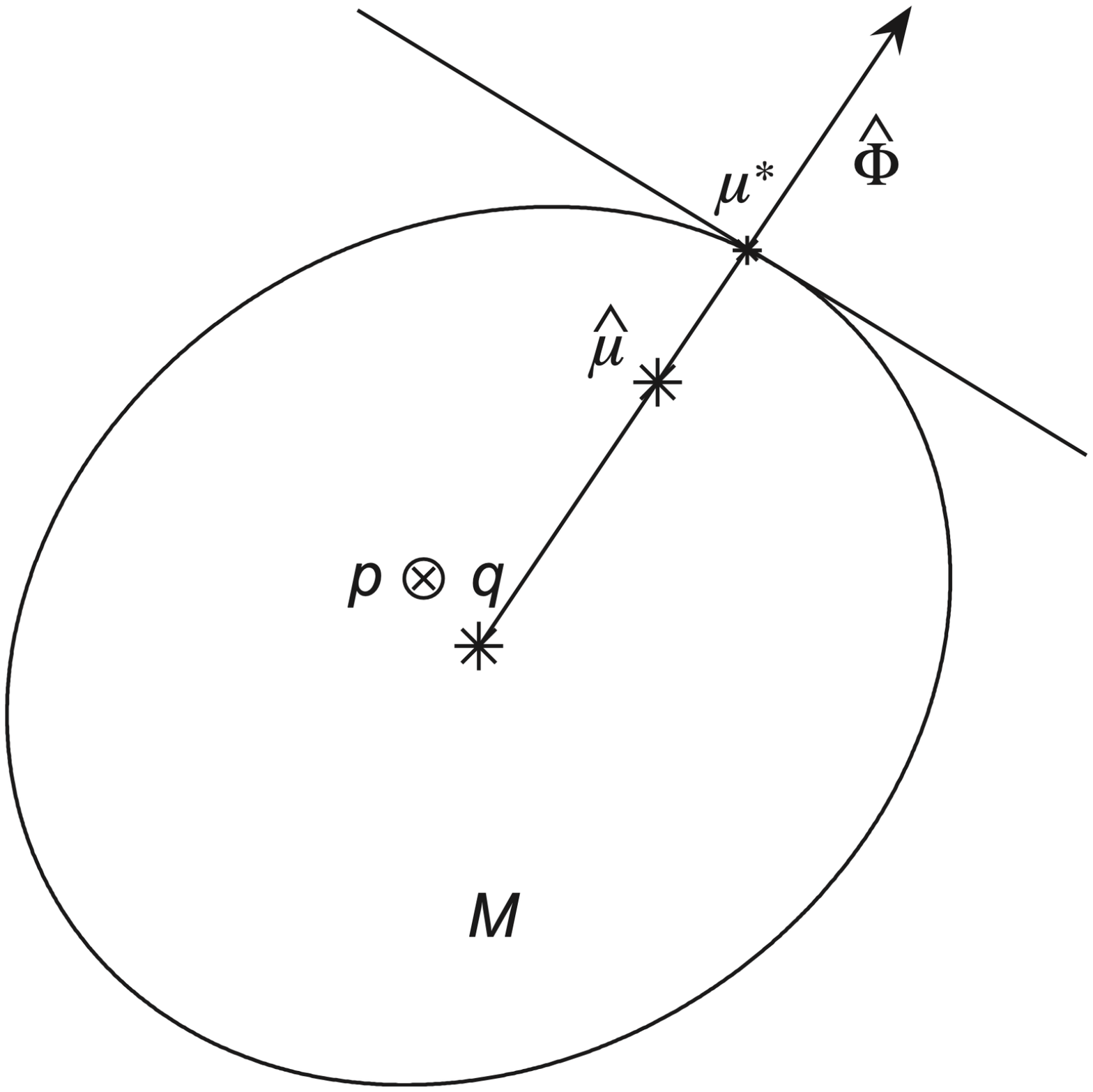}
\caption{Geometric illustration of Example \protect\ref{ex:gauge}. $\hat{\protect\mu}$ is not rationalizable, but it is associated to some proximate $\protect\mu ^{\ast }$ on the boundary of $\mathcal{M}$, which is itself rationalized by $\hat{\Phi}$.}
\label{Fig2}
\end{center}
\end{figure}

\begin{example}
\label{ex:gauge}Recall the definition $\left( p\otimes q\right)
_{xy}=p_{x}p_{y}$, and remember that $p\otimes q$ is never on the frontier
of $\mathcal{M}$, hence never rationalizable. When $\hat{\mu}$ is not
rationalizable either, one may consider the smallest $t$ such that $p\otimes
q+t\left( \hat{\mu}-p\otimes q\right) $ is rationalizable. This number
exists and is finite because the halfline which starts from $p\otimes q$
through $\hat{\mu}$ must cross the frontier of $\mathcal{M}$, which is a
convex and compact set. Letting $t^{\ast }$ be the corresponding value of $t$%
, and $\mu ^{\ast }=p\otimes q+t^{\ast }\left( \hat{\mu}-p\otimes q\right) $%
, there exists by definition an element $\hat{\Phi}\in \mathbf{S}\backslash
\left\{ 0\right\} $ such that $\mu ^{\ast }\in \partial \mathcal{W}%
_{0}\left( \hat{\Phi}\right) $, where $\mathcal{W}_{0}$ is as in (\ref{SWF0}%
). Note that if $\hat{\mu}$ is rationalizable, then $t^{\ast }=1$ and $\mu
^{\ast }=\mu $. See Figure \ref{Fig2}.

This construction can be expressed in terms of $\mathcal{I}$. Letting%
\begin{eqnarray}
\mathcal{I}\left( \hat{\mu}\right) &=&-\max_{t\geq 1}\left\{ t:p\otimes
q+t\left( \hat{\mu}-p\otimes q\right) \in \mathcal{M}\right\} \text{ if }%
\hat{\mu}\in \mathcal{M}  \label{I1} \\
&=&+\infty \text{ else}
\end{eqnarray}%
so that $\mathcal{I}\left( \hat{\mu}\right) $ can be formulated as a max-min
problem, that is, for $\hat{\mu}\in \mathcal{M}$,
\begin{equation*}
\mathcal{I}\left( \hat{\mu}\right) =-\max_{t\geq 1}\min_{\Phi \in \mathbf{S}%
}\left\{ t+\mathcal{W}_{0}\left( \Phi \right) -\left\langle \Phi ,p\otimes
q+t\left( \hat{\mu}-p\otimes q\right) \right\rangle \right\} .
\end{equation*}

Because the objective function is convex in $\Phi $ and linear in $t$, this
problem has a saddlepoint which will be denoted $\left( \Phi ^{\ast
},t^{\ast }\right) $. Let $\mu ^{\ast }=p\otimes q+t^{\ast }\left( \hat{\mu}%
-p\otimes q\right) $. By optimality with respect to $\Phi $, $\mu ^{\ast
}\in \partial \mathcal{W}_{0}\left( \Phi ^{\ast }\right) $, thus $\Phi
^{\ast }$ rationalizes $\mu ^{\ast }$. By the envelope theorem
\begin{equation*}
t^{\ast }\Phi ^{\ast }=\nabla \mathcal{I}\left( \hat{\mu}\right) ,
\end{equation*}%
thus we take
\begin{equation*}
\hat{\Phi}=t^{\ast }\Phi ^{\ast }.
\end{equation*}

and $\mu ^{\ast }$ is the matching which is on the halfline which starts
from $p\otimes q$ through $\hat{\mu}$ and which is rationalizable.
\end{example}

\bigskip

\begin{example}
\label{ex:CS}In the Choo and Siow (2005) model, the surplus function is $%
\Phi _{ij}=\Phi \left( x,y\right) +\varepsilon _{iy}+\eta _{jx}$ where $%
\varepsilon _{iy}$ and $\eta _{jx}$ are iid extreme value type I random
variables. Choo and Siow use this model nonparametrically identifies $\Phi $%
. Galichon and Salani\'{e} (2010) show that this model leads to the
following specification of $\mathcal{I}$:
\begin{eqnarray}
\mathcal{I}\left( \mu \right) &=&\sum_{xy}\mu _{xy}\log \mu _{xy}\text{ if }%
\mu \in \mathcal{M}  \label{I2} \\
&=&+\infty \text{ else.}  \notag
\end{eqnarray}
\end{example}

\bigskip

\begin{example}
\label{ex:GS}Galichon and Salani\'{e} (2012) argue that the model of Choo
and Siow actually extends in the case where the matching surplus function in
the presence of heterogeneities between man $i$ of type $x$ and woman $j$ of
type $y$ is $\Phi _{ij}=\Phi \left( x,y\right) +\varepsilon _{ixy}+\eta
_{jxy}$, and letting $G_{x}\left( U\right) =\mathbb{E}\left[ \max_{y}\left(
U_{xy}+\varepsilon _{ixy}\right) \right] $ and $H_{y}\left( V\right) =%
\mathbb{E}\left[ \max_{x}\left( V_{xy}+\eta _{jxy}\right) \right] $ be the
ex-ante indirect utilities of respectively the man of type $x$ and the woman
of type $y$, and letting $G^{\ast }$ and $H^{\ast }$ their respective convex
conjugate transforms, that is%
\begin{eqnarray*}
G_{x}^{\ast }\left( \mu _{.|x}\right) &=&\sup_{U_{xy}}\{\sum_{y}\mu
_{y|x}U_{xy}-G_{x}\left( U\right) \}\text{ if }\sum_{y}\mu _{y|x}=1 \\
&=&+\infty \text{ else,}
\end{eqnarray*}%
and%
\begin{eqnarray*}
H_{y}^{\ast }\left( \mu _{.|y}\right) &=&\sup_{V_{xy}}\{\sum_{x}\mu
_{x|y}V_{xy}-H_{y}\left( V\right) \}\text{ if }\sum_{x}\mu _{x|y}=1 \\
&=&+\infty \text{ else.}
\end{eqnarray*}

Then $\mathcal{I}\left( \mu \right) $ is given by%
\begin{equation}
\mathcal{I}\left( \mu \right) =\sum_{x}p_{x}G_{x}^{\ast }\left( \mu
_{.|x}\right) +\sum_{y}q_{y}H_{y}^{\ast }\left( \mu _{.|y}\right) .
\label{I3}
\end{equation}%
which coincides with (\ref{I2}) in the case studied by Choo and Siow, hence
the term \textquotedblleft generalized entropy\textquotedblright . As an
important consequence, this paves the way to the continuous generalization
of the Choo and Siow model. See Dupuy and Galichon (2012), and Bojilov and
Galichon (2013).
\end{example}

\bigskip

\begin{example}
Applying this setting, Galichon and Salani\'{e} (2012, Example 3) assume
that $\mathcal{X}$ and $\mathcal{Y}$ are finite subsets of $\mathbb{R}$, and
that $\varepsilon _{ixy}=e_{i}y$ while $\eta _{jxy}=f_{j}x$ where $e_{i}$
and $f_{j}$ are drawn from $\mathcal{U}\left( \left[ 0,1\right] \right) $
distributions. In this case the utility shocks are perfectly correlated
across alternatives, in sharp contrast with Example 1, where they are
independent. Then, letting $Q_{Y|X=x}^{\mu }$ be the conditional quantile of
$Y$ conditional on $X=x$ under distribution $\mu $, one has%
\begin{equation*}
\mathcal{I}\left( \mu \right) =\sum_{x}p_{x}\int_{0}^{1}Q_{Y|X=x}^{\mu
}\left( t\right) tdt+\sum_{y}q_{y}\int_{0}^{1}Q_{X|Y=y}^{\mu }\left(
t\right) tdt.
\end{equation*}
\end{example}

\bigskip

{\small $^{\S }$}\textit{\ Sciences Po Paris, Department of Economics,
Address: 28 rue des Saint-P\`{e}res, 75007 Paris, France. E-mail:
alfred.galichon@sciences-po.fr.} 

\appendix

\section*{Facts from Convex Analysis}

The definitions below are included for completeness and the reader is
referred to Ekeland and Temam (1976) for a thorough exposition of the topic.

Take any set $Y\subset \mathbb{R}^{d}$; then the \emph{convex hull\/} of $Y$
is the set of points in $\mathbb{R}^{d}$ that are convex combinations of
points in $Y$. We usually focus on its closure, the closed convex hull,
denoted $cch\left( Y\right) $.

The \emph{support function} $S_{Y}$ of $Y$ is defined as
\begin{equation*}
S_{Y}\left( x\right) =\sup_{y\in Y}x\cdot y
\end{equation*}%
for any $x$ in $Y$, where $x\cdot y$ denotes the standard scalar product. It
is a convex function, and it is homogeneous of degree one. Moreover, $%
S_{Y}=S_{\mbox{cch}\left( Y\right) }$ where $\mbox{cch}\left( Y\right) $ is
the closed convex hull of $Y$, and $\partial S_{Y}\left( 0\right) =\mbox{cch}%
\left( Y\right) $.

A point in $Y$ is an \emph{boundary point} if it belongs in the closure of $%
Y $, but not in its interior.

Now let $u$ be a convex, continuous function defined on $\mathbb{R}^{d}$.
Then the gradient $\nabla u$ of $u$ is well-defined almost everywhere and
locally bounded. If $u$ is differentiable at $x$, then
\begin{equation*}
u\left( x^{\prime }\right) \geq u\left( x\right) +\nabla u\left( x\right)
\cdot (x^{\prime }-x)
\end{equation*}%
for all $x^{\prime }\in \mathbb{R}^{d}$. Moreover, if $u$ is also
differentiable at $x^{\prime }$, then
\begin{equation*}
\left( \nabla u\left( x\right) -\nabla u\left( x^{\prime }\right) \right)
\cdot \left( x-x^{\prime }\right) \geq 0.
\end{equation*}%
When $u$ is not differentiable in $x$, it is still \emph{\
subdifferentiable\/} in the following sense. We define $\partial u\left(
x\right) $ as%
\begin{equation*}
\partial u\left( x\right) =\left\{ y\in \mathbb{R}^{d}:\forall x^{\prime
}\in \mathbb{R}^{d},u\left( x^{\prime }\right) \geq u\left( x\right) +y\cdot
(x^{\prime }-x)\right\} .
\end{equation*}%
Then $\partial u\left( x\right) $ is not empty, and it reduces to a single
element if and only if $u$ is differentiable at $x$; in that case $\partial
u\left( x\right) =\left\{ \nabla u\left( x\right) \right\} $.

Given a convex function $u$ defined on a convex subset of $\mathbb{R}^{d}$,
one defines its \emph{convex conjugate} as
\begin{equation*}
u^{\ast }\left( y\right) =\sup_{x\in \mathbb{R}^{d}}\left\{ x\cdot y-u\left(
x\right) \right\} .
\end{equation*}

One has $y\in \partial u\left( x\right) $ if and only if $x\in \partial
u^{\ast }\left( y\right) $ if and only if $u\left( x\right) +u^{\ast }\left(
y\right) =x\cdot y$.

\end{document}